\documentclass[
 ,draft            
  ]
{aipproc}  

\layoutstyle{6x9} 

\newcommand{\beq}{\begin{equation}}
\newcommand{\eeq}{\end{equation}}

\newcommand{\di}{\displaystyle}

\newcommand{\si}{\sigma}

\newcommand{\nubar}[0]{\overline{\nu}_\mu}

\begin{document}

\title{Extraction of Neutrino Flux with the Low $\nu$ Method at MiniBooNE Energies}



\author{A. Bodek}{
  address={Department of Physics and Astronomy,
             University of Rochester,
             Rochester, New York, 14618,  USA}
}

\author{U. Sarica}{
  address={Department of Physics and Astronomy,
             University of Rochester,
             Rochester, New York, 14618,  USA}
             }
             
\author{K. S. Kuzmin}{
  address={Bogoliubov Lab. of Theoretical Physics, Joint Inst. for Nuclear Research, 141980, Dubna, Russia}
 ,altaddress={Institute for Theoretical  and Experimental Physics, 117218 Moscow, Russia} 
}

\author{V. A. Naumov}{
  address={Bogoliubov  Lab. of Theoretical Physics, Joint Inst.  for Nuclear Research, 141980, Dubna, Russia}
}


\begin{abstract}
We describe the application of the `low-$\nu$' method to the extraction of the neutrino
flux at MiniBooNE energies. As an example, we extract the relative energy dependence of the flux 
 from published MiniBooNE  quasielastic scattering cross sections with 
$\nu < 0.2$ GeV and $\nu < 0.1$ GeV (here $\nu$ is the energy transfer to the target). We find that the flux extracted from the  `low-$\nu$' cross sections is consistent with the nominal flux used by MiniBooNE. We
 fit the MiniBooNE cross sections over the entire kinematic range to various parametrizations of the axial form factor. 
We find that if the
overall normalization of the fit is allowed to float within the normalization errors, the extracted values of the axial vector mass are independent of the flux.   Within the Fermi gas
model, the  $Q^2$ distribution
of the MiniBooNE data is described by a standard dipole form factor with  $M_A=1.41\pm0.04$ GeV.  If nuclear transverse enhancement in the  vector form factors is accounted for,  the data  are best
 fit with a modified dipole form factor with $M_A=1.10\pm 0.03$ GeV. 

\end{abstract}

\maketitle



In a previous communication \citep{lownu} we present the application of the `low-$\nu$' method
to the extraction of neutrino ($\nu_\mu$)  flux for energies ($E_\nu$)  as low as 0.7 GeV.  In this paper we extend the technique
to $E_\nu$ as low as  0.4 GeV and extract the relative energy dependence of the $\nu_\mu$ flux for the MiniBooNE experiment as an 
example.

The charged current $\nu_\mu$ ($\nubar$) differential cross section  can be written in terms of the square of the four momentum ($Q^2$) and  energy transfer ($\nu$) to the target nucleus.
At low-$\nu$,   if we integrate the
cross section  from $\nu_{min}\approx 0$ up to  $\nu$= $\nu_{cut}$ (where $\nu_{cut}$ is small), we can write the `low-$\nu$' cross section\citep{lownu}  in terms of  an energy independent term which is proportional to the  structure function ${\cal W}_2$,  and small energy dependent corrections which are proportional to $\nu/E$, or $m_\mu^2/E^2$ where $m-\mu$ is the mass of the muon.

$$ \sigma_{\nu cut}(E_\nu)  = \displaystyle\int_{\nu_{min (E_\nu)}}^{\nu_{cut}}   \di\frac{d^2\si}{dQ^2 d\nu}dQ^2 d\nu 
= \sigma_{{W}_2} + \sigma_{2}+\sigma_{1}\pm\sigma_{3}+\sigma_{4} + \sigma_{5}$$
$$\sigma_{{W}_2} = C  \displaystyle\int_{\nu_{min (E_\nu)}}^{\nu_{cut}}  {\cal W}_2 ~d\nu;~~~~~~~~~~
~~~~~~~~\sigma_{2} =C \displaystyle\int_{\nu_{min (E_\nu)}}^{\nu_{cut}} 
  \left[ -\frac{\nu}{E_\nu} -\frac{Q^2+m_\mu^2}{4E_\nu^2}  \right]{\cal W}_2 ~d\nu$$
$$ \sigma_{1}=C \displaystyle\int_{\nu_{min (E_\nu)}}^{\nu_{cut}}
\frac{(Q^2+m_\mu^2)}{2E_\nu^2} {\cal W}_1 ~d\nu;~~~
\sigma_{3}=C \displaystyle\int_{\nu_{min (E_\nu)}}^{\nu_{cut}} 
\left[ \frac{Q^2}{2ME_\nu}-\frac{\nu}{4E_\nu}  \frac { Q^2+m_\mu^2}{ME_\nu}  \right]   {\cal W}_3 ~d\nu,$$ 
where $\sigma_{4}$  and $\sigma_{5}$ are negligible\citep{lownu}, and  $M$ is the nucleon mass.
The  uncertainties in the modeling of the small energy dependent correction terms are small, thus we
can extract the relative $\nu_\mu$ flux from  the  number
of  low-$\nu$ events at each $E_\nu$ bin.
\begin{figure}[h]
 \includegraphics[height=0.20\textheight,width=0.33\textheight]{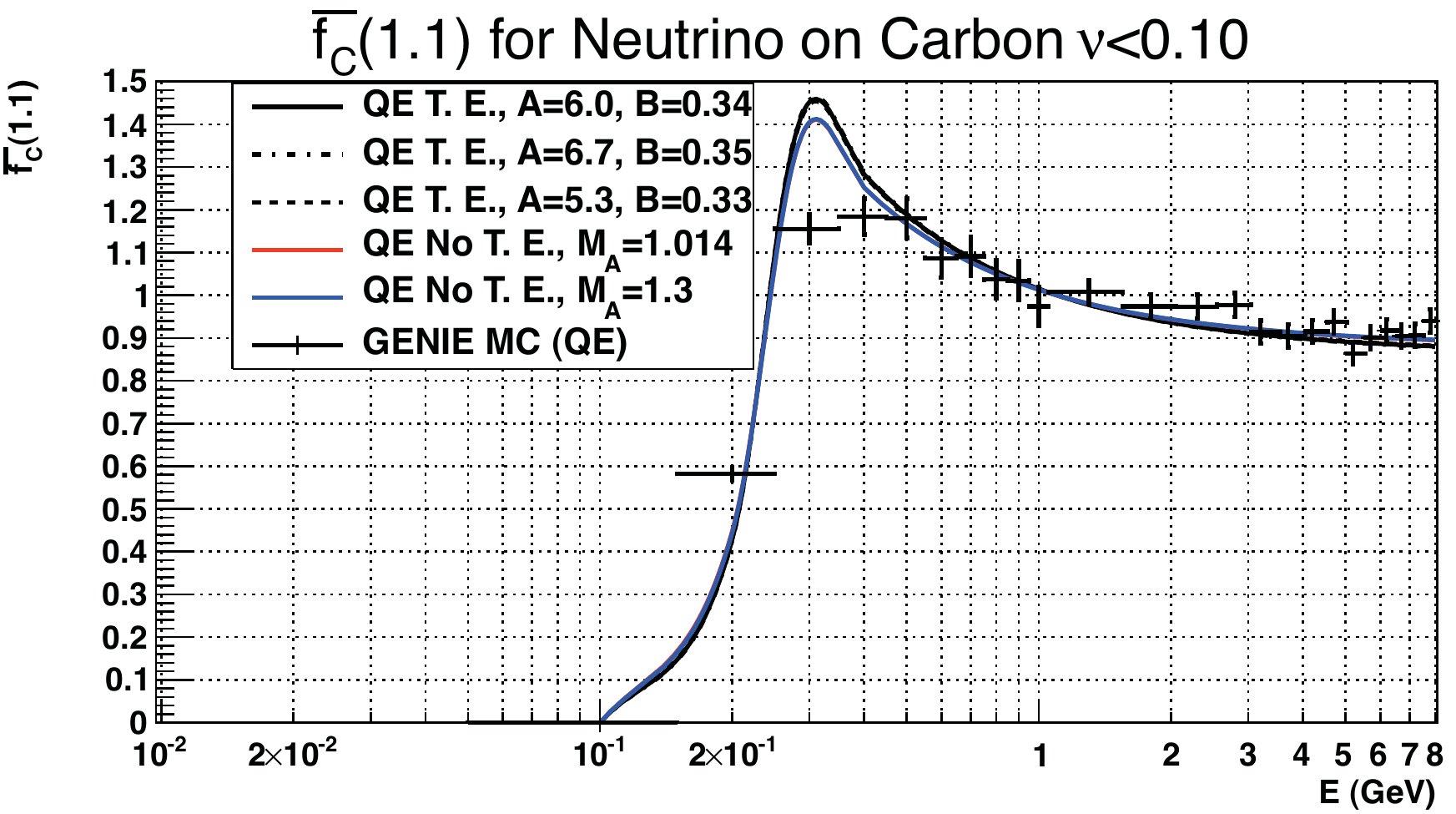}
  \includegraphics[height=0.20\textheight,width=0.33\textheight]{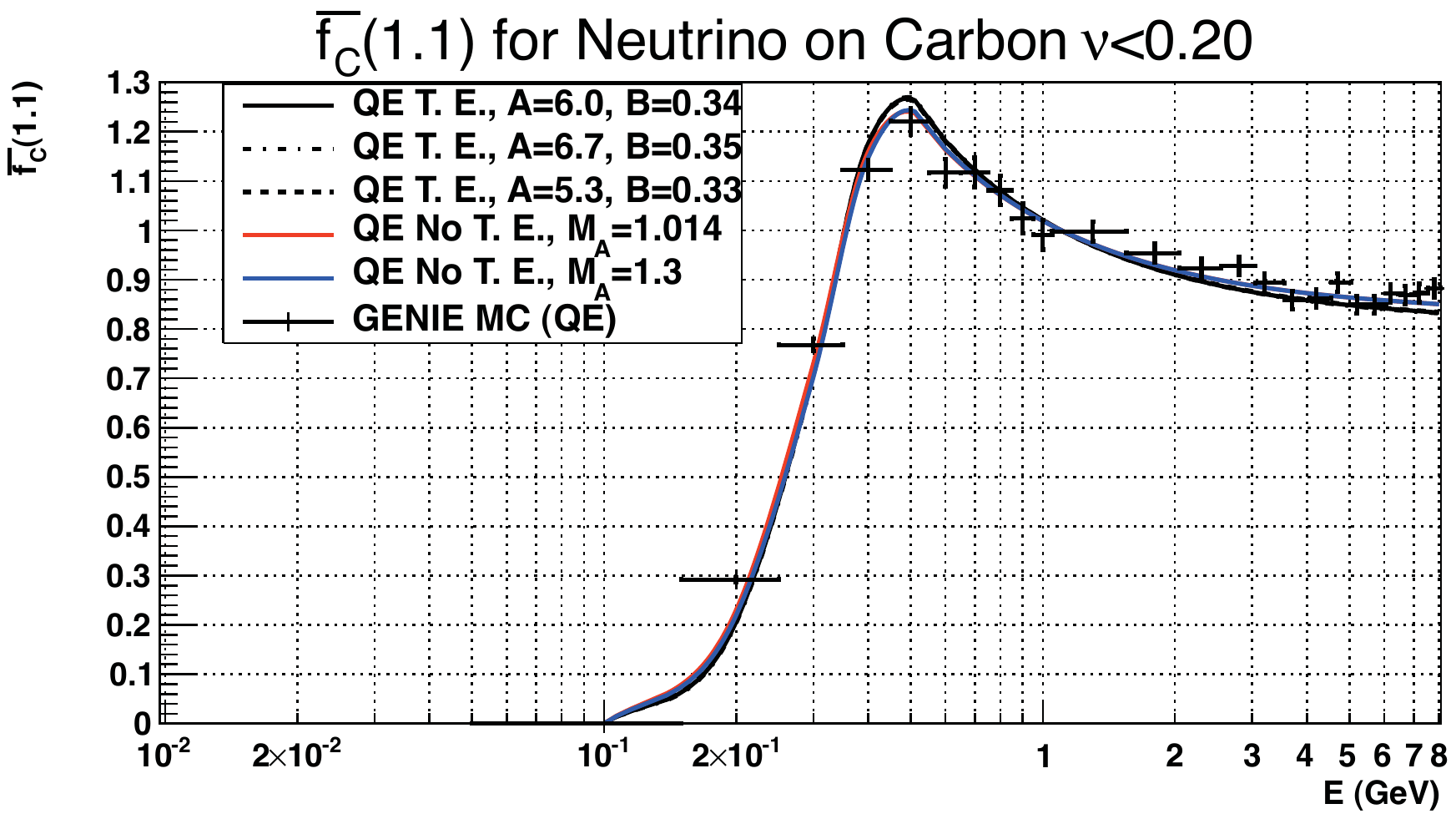}
   \label{fig-nu}
  \caption{ The ratio of the neutrino  `low-$\nu$' QE cross section (as a function of $E_\nu$) to the `low-$\nu$' QE cross section
  at $E_\nu=1.1$ GeV  for $\nu<0.1$ GeV (left) and $\nu<0.2$ GeV (right). }
\end{figure} 
\begin{figure}[h]
 \includegraphics[height=0.20\textheight,width=0.33\textheight]{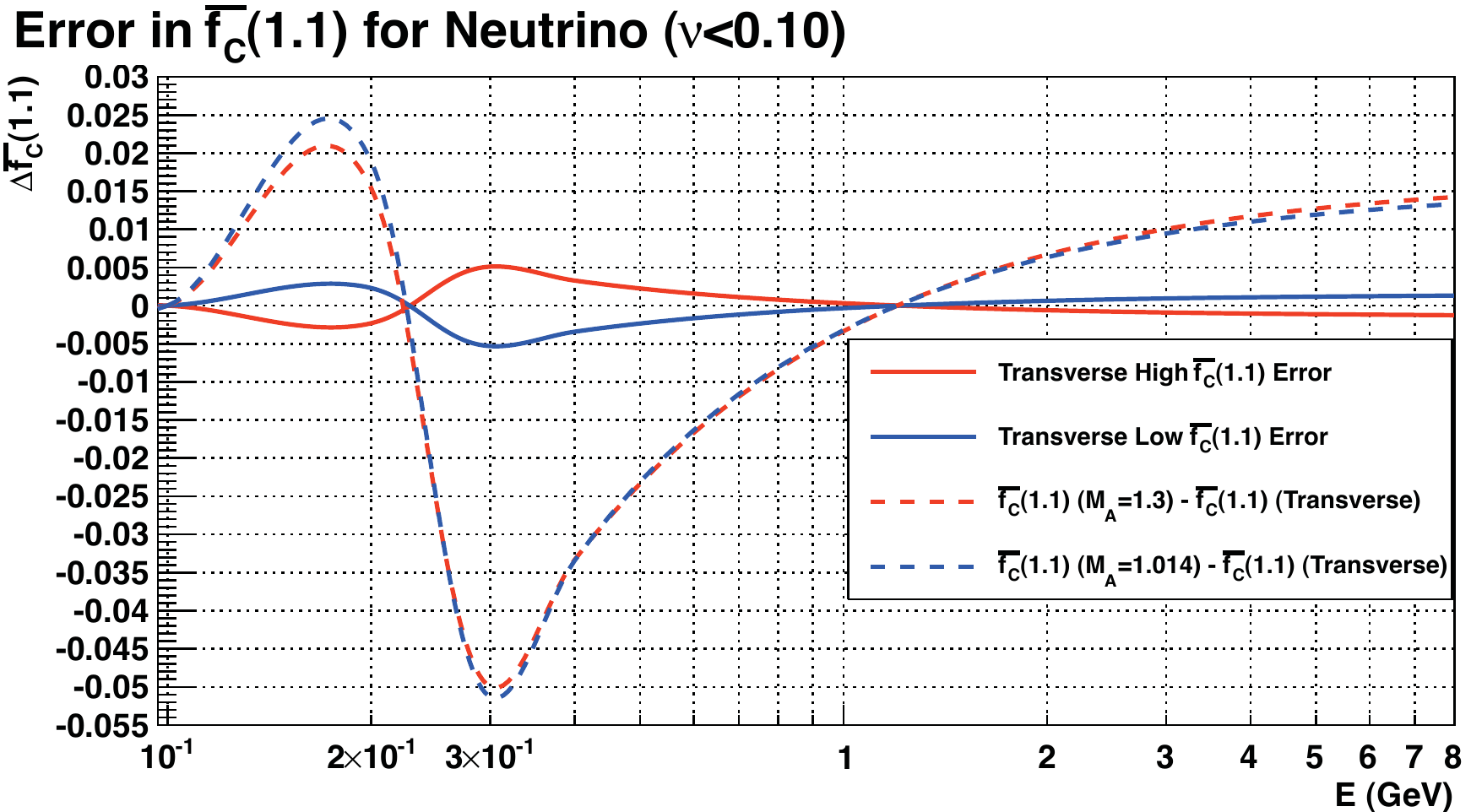}
  \includegraphics[height=0.20\textheight,width=0.33\textheight]{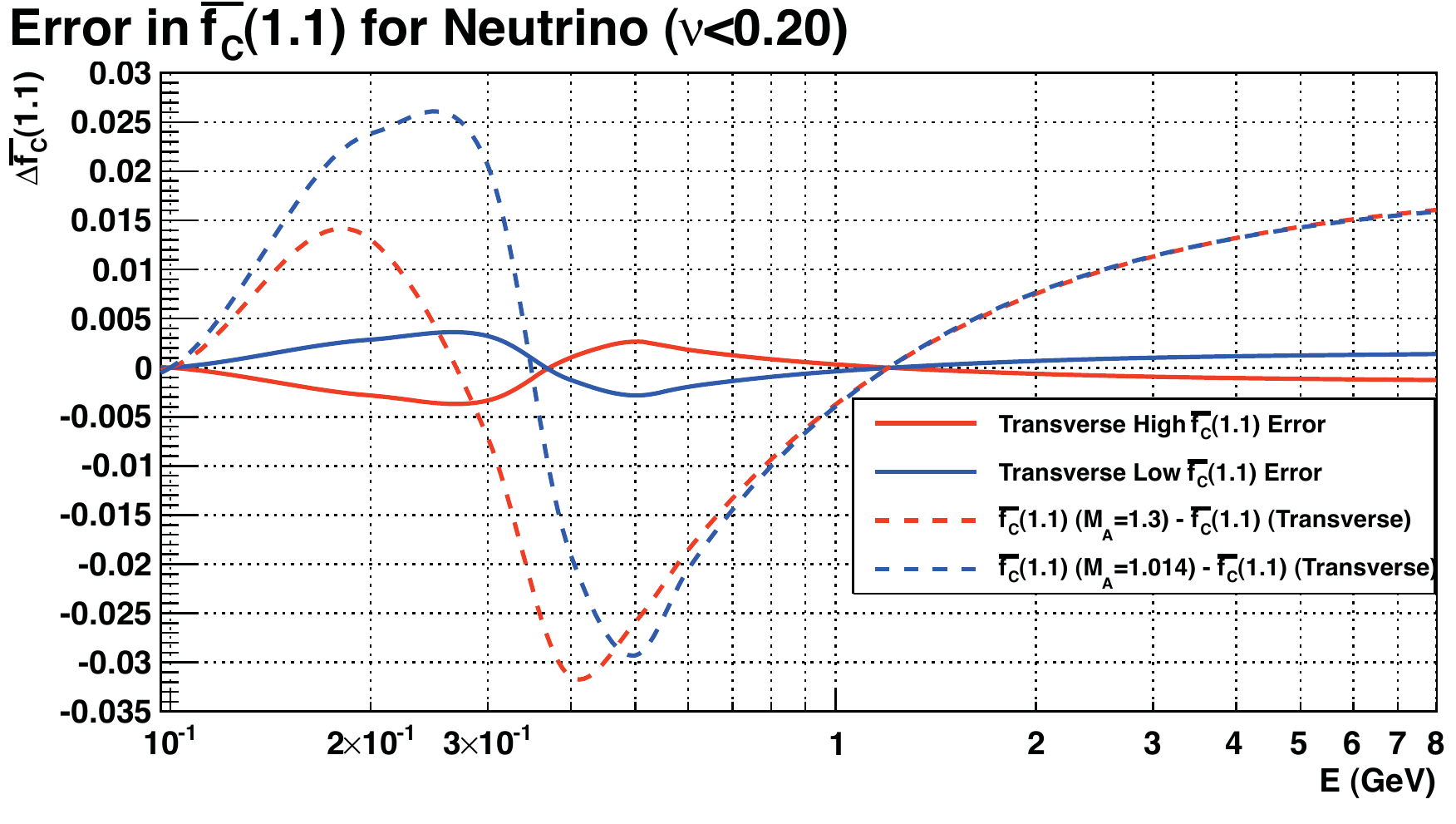}
   \label{fig-nuerr}
  \caption{ The  model uncertainty in the  ratio of the neutrino  `low-$\nu$' QE cross section (as a function of $E_\nu$) to the `low-$\nu$' QE cross section
  at $E_\nu=1.1$ GeV  for $\nu<0.1$ GeV (left) and $\nu<0.2$ GeV (right). 
  }
\end{figure}

If we use the  MINOS criteria that the fraction
of events in the `low-$\nu$'  flux sample is lower than 60\% of the total  number
of events in each $E_{\nu,\nubar}$ bin,  we find that for neutrinos  we can use events with $\nu<0.1$ GeV
to extract the relative flux for  $E_\nu>0.4$ GeV, and events with  
$\nu<0.2$ GeV for $E_\nu>0.7$ GeV.  For these $\nu$ cuts, the cross section is dominated
by quasielastic (QE) scattering. 
The flux extracted with the `low-$\nu$' method is only a relative flux as a function of energy.
 It must be normalized at some energy.  In this paper, we present the flux relative to the flux  
 at $E_\nu$= 1.1  GeV.  In our calculation of QE cross sections we use
 BBBA2007 electromagnetic form factors\citep{BBB}.
 
\begin{figure}[h]
 \includegraphics[height=0.20\textheight,width=0.35\textheight]{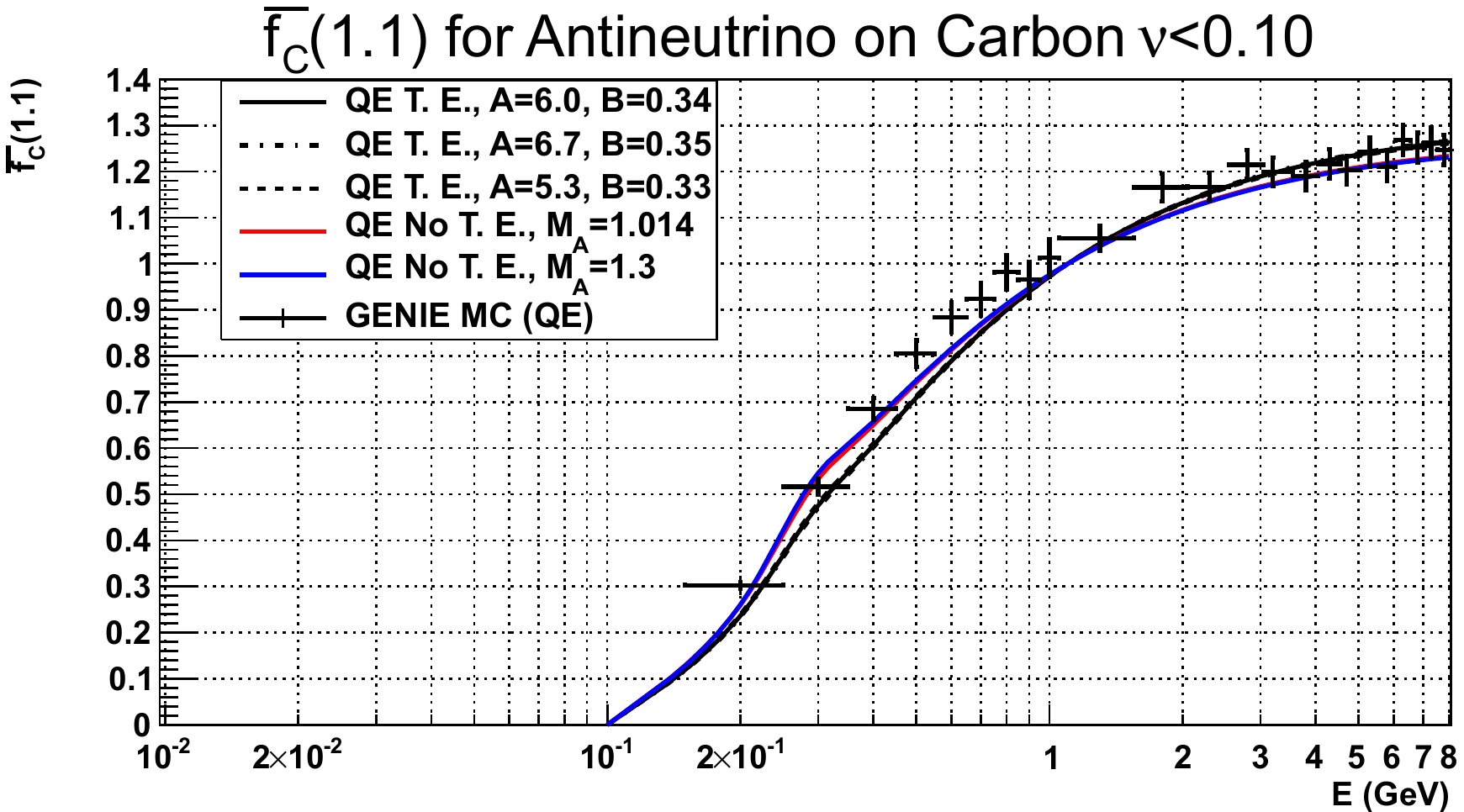}
  \includegraphics[height=0.20\textheight,width=0.35\textheight]{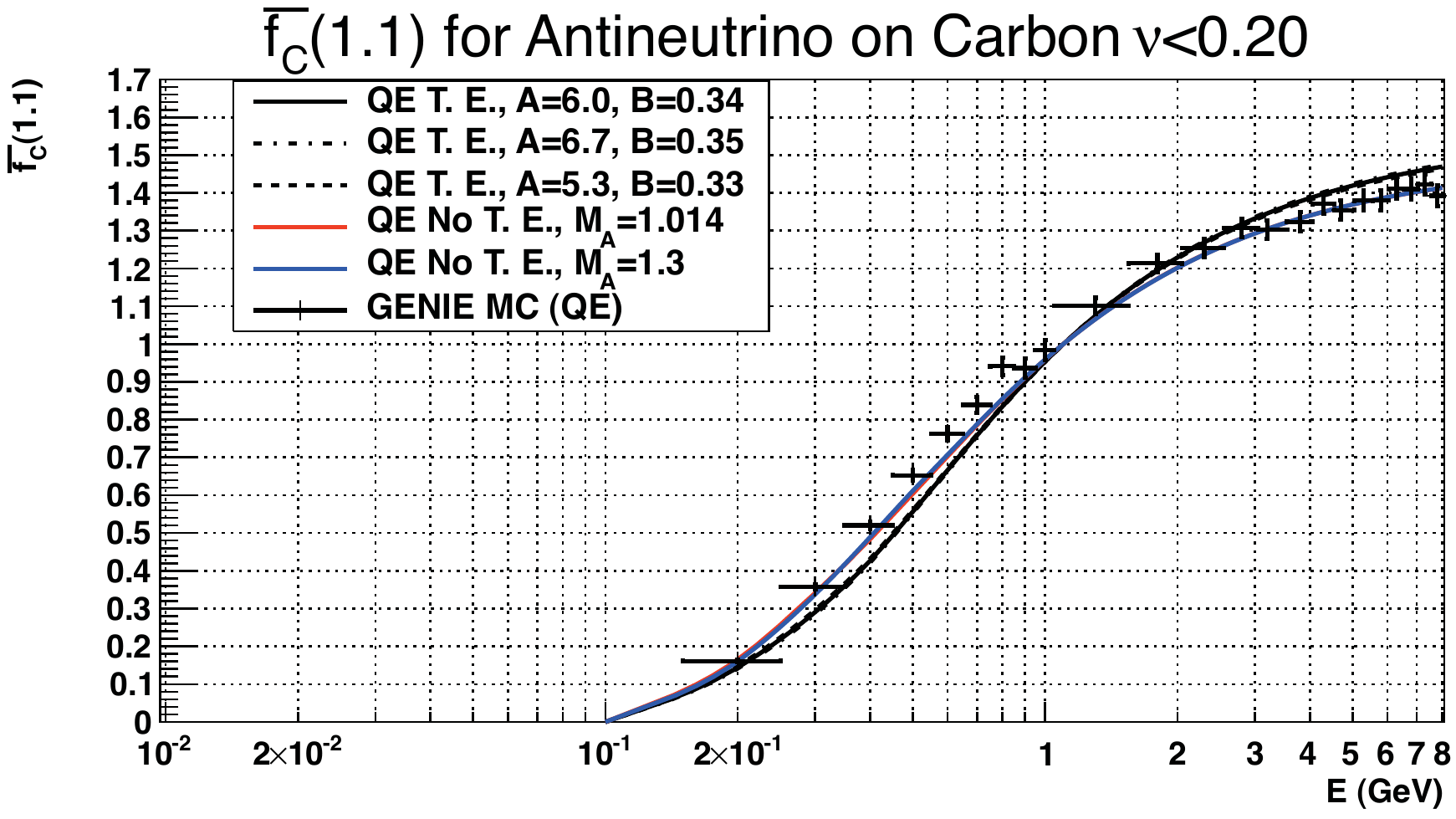}
   \label{fig-nubar}
  \caption{ The ratio of the $\nubar$  `low-$\nu$' QE cross section (as a function of $E_{\nubar}$) to the `low-$\nu$' QE cross section at $E_{\nubar}=1.1$ GeV  for $\nu<0.1$ GeV (left) and $\nu<0.2$ GeV (right). }
\end{figure}

\begin{figure}[h]
 \includegraphics[height=0.20\textheight,width=0.35\textheight]{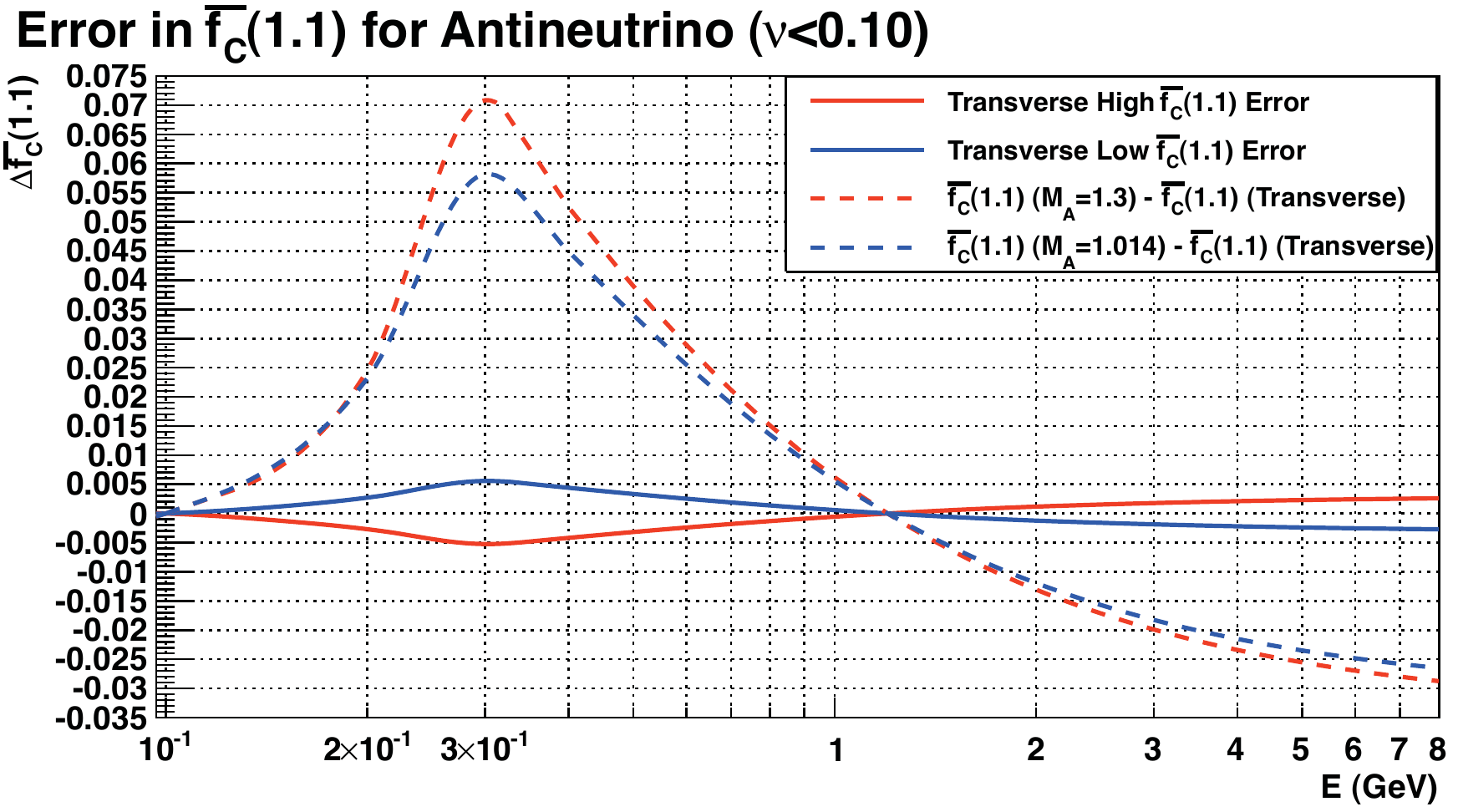}
  \includegraphics[height=0.20\textheight,width=0.35\textheight]{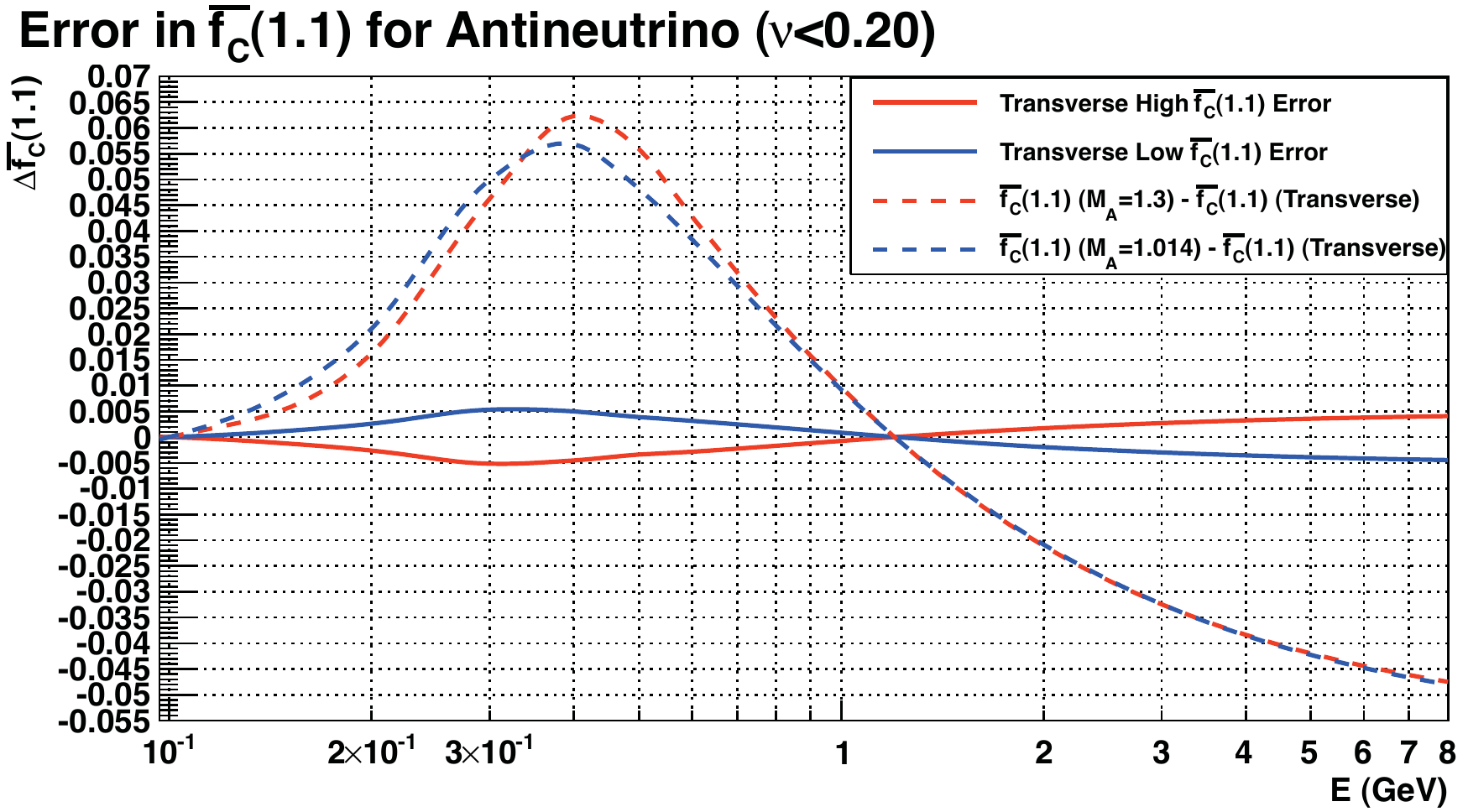}
   \label{fig-nubarerr}
  \caption{ The  model uncertainty in the ratio of the $\nubar$   `low-$\nu$' QE cross section (as a function of $E_{\nubar}$) to the `low-$\nu$' QE cross section
  at $E_{\nubar}=1.1$ GeV  for $\nu<0.$ GeV (left) and $\nu<0.2$ GeV (right).
}
\end{figure}

 Figures \ref{fig-nu} ($\nu_\mu$) and \ref{fig-nubar} ($\nubar$)  show the ratio of the `low-$\nu$' QE cross section (as a function of $E_{\nu,\nubar}$)  to the `low-$\nu$' cross section 
  at $E_{\nu,\nubar}=1.1$ GeV  for $\nu<0.1$ GeV (left) and $\nu<0.2$ GeV (right) for various models. 
  The data points are from the GENIE MC generator for a carbon target assuming a  Fermi gas model and a dipole form for the axial form factor with   $M_A$ = 0.99 GeV\citep{Kuzmin}.  The ratio is independent
  of the value $M_A$ as illustrated by the fact that 
   prediction of this ratio for a  dipole axial vector mass   $M_A$ = 1.014 GeV (solid red line) and $M_A$ = 1.3 GeV (solid blue line) are the same.
     Also shown are the changes in the prediction when we include  nuclear enhancement in the transverse vector form factors\citep{TE} (TE) (shown as the solid  black line). For $E_{\nu}>0.4$ GeV  and $\nu<$0.1 and for
     $E_{\nu}>0.7$ GeV and $\nu<$0.2  the ratio is approximately constant.  The uncertainties in the  modeling of the energy dependence of
     this ratio are small as shown in figures  \ref{fig-nuerr} and \ref{fig-nubarerr}.    
As an example, we extract the relative energy dependence of the flux of the Fermilab booster neutrino beam from  MiniBooNE data.
 The MiniBooNE experiment published\citep{MiniBooNE}  flux weighted  double differential cross sections for QE  neutrino scattering in bins of  final state muon kinetic energy $T_\mu$ ($E_\mu=T_\mu +m_\mu$)  and muon angle ($\cos\theta_{\mu}$).
  We extract the central value of
$\nu^{QE}=E_\nu^{QE}-E_\mu$ for each  ($T_\mu$, $\cos\theta_{\mu}$) bin using
 \vspace{-0.1in}
\begin{eqnarray}
   E_\nu^{QE}
&=&
\frac{2(M_n^{\prime})E_\mu-((M_n^{\prime})^2+m_\mu^2-M_p^2)}
{2\cdot[(M_n^{\prime})-E_\mu+\sqrt{E_\mu^2-m_\mu^2}\cos\theta_\mu]}.
\end{eqnarray}
where $M_n$ and $M_p$
are the neutron and proton mass, and  $M_n^{\prime}=M_n-E_B$ ($E_B= 34$~MeV).

\begin{figure} [ht]
 \includegraphics[height=0.25\textheight,width=0.37\textheight]{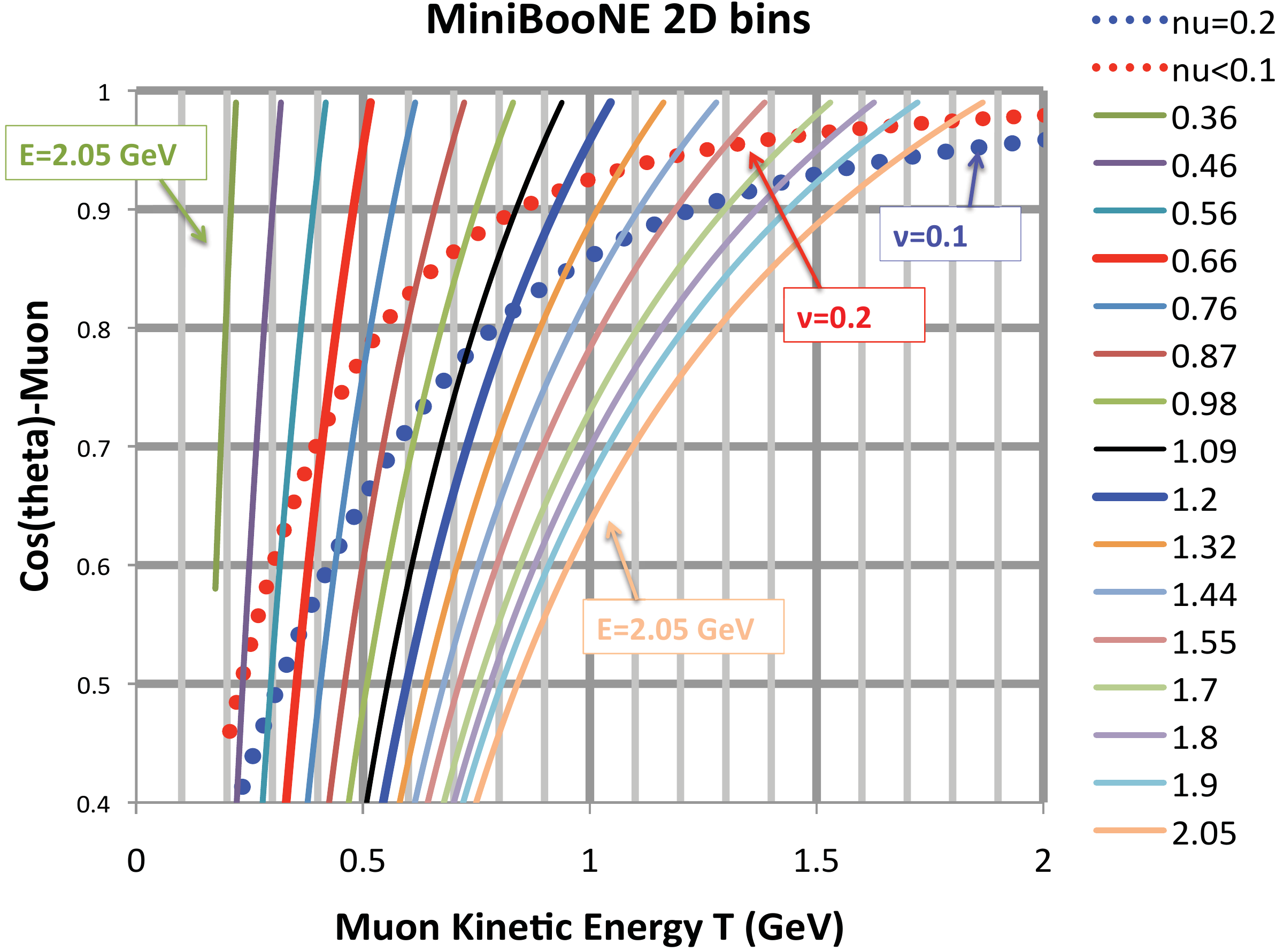}
  \includegraphics[height=0.24\textheight,width=0.35\textheight]{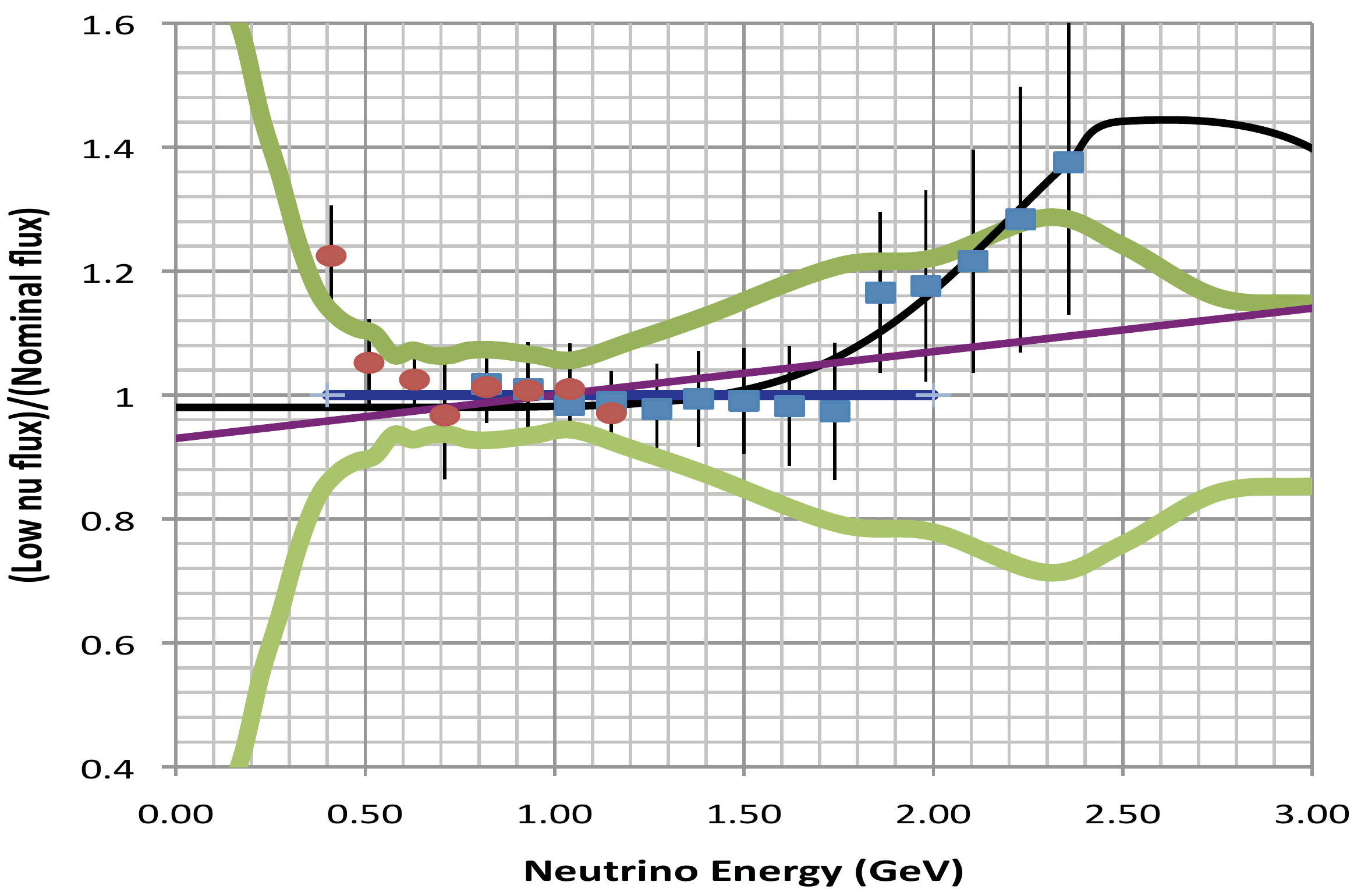}
   \label{fig-flux}
  \caption{Left:  The MiniBooNE QE cross section bins of 0.1
  in $\cos\theta_\mu$, and 0.1 GeV in $T_\mu$. The solid lines are
  lines of constant $E_\nu^{QE}$.  The  blue and red dotted lines are
  $\nu^{QE}<0.2$ GeV, and $\nu^{QE}<0.1$ GeV, respectively.  Right:    The relative neutrino flux extracted
 from $\nu^{QE}<0.2$ GeV cross sections (blue squares) and  $\nu^{QE}<0.1$ GeV (red squares) shown with  statistical errors only. The black (flux A) and purple (flux B)  lines are possible deviations (which are consistent with the `low-$\nu$' flux) from the central values of the published flux.  The green line
  is the quoted systematic uncertainty in the nominal  MiniBoonNE flux.
}
\end{figure}

The left side of Fig.~\ref{fig-flux} shows the MiniBooNE bins of 0.1
GeV  in $T_\mu$ and 0.1 
  in $\cos\theta_\mu$.  The solid lines are
  lines of constant $E_\nu^{QE}$.  The  blue and red dotted lines are
  $\nu^{QE}<0.2$ GeV, and $\nu^{QE}<0.1$ GeV, respectively.  
\begin{figure} [ht]
 \includegraphics[height=0.25\textheight,width=0.70\textheight]{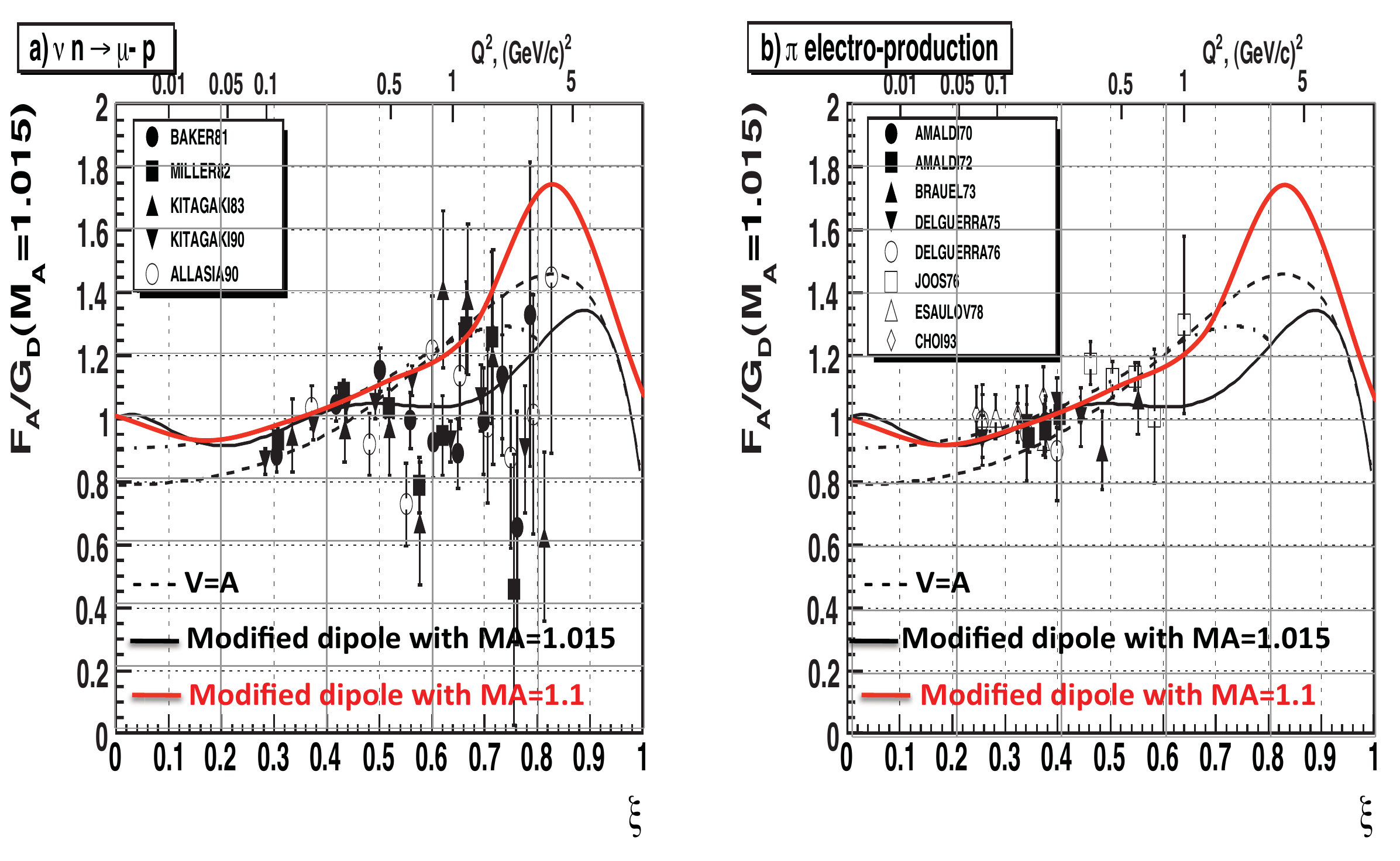}
     \label{fig-fa}
  \caption{ $F_A(Q^2)$ measurements on free nucleons  (a) $F_A(Q^2)$ re-extracted from neutrino-deuterium
 data divided by $G_D^{A}(Q^2)$ (with $M_A=1.015$ GeV). (b)  
$F_A(Q^2)$ from pion electroproduction
  divided by $G_D^{A}(Q^2)$  corrected for
  for hadronic effects\citep{pion}.
  Solid black line - duality based modified dipole fit with $M_A=1.015 GeV$\citep{BBB}. 
  Short-dashed line - $F_A(Q^2)_{A2=V2}$. 
  Dashed-dot line - constituent quark model\citep{quark}. Solid red line - duality based modified dipole with $M_A=1.10$ GeV, which is our best fit to the MiniBooNE data on Carbon (accounting for Transverse Enhancement).  
}
\end{figure}
  We extract the `low-$\nu$' flux from the MiniBooNE data as follows. Using the published MiniBooNE flux,
  we first fit the flux-weighted doubly differential cross section to three models. The parameters
  which are allowed to float within the models are the overall normalization and the axial vector mass
  $M_A$. The first model is  a Fermi gas model with BBBA2007 electromagnetic
  form factors and a dipole form for the axial form factor. The second model includes Transverse
  Enhancement for the vector form factors\citep{TE} (BBC-TE) and a dipole form for the axial form factor.  The third  model  includes TE
for the vector form factors\citep{TE} (BBC-TE) and assumes a modified dipole form for the axial form factor
  as given in ref.~\citep{BBB}.  The modification to the dipole form factor are from a
  fit\citep{BBB}  to all neutrino scattering data and pion electroproduction on free (H and D targets) nucleons 
  as shown in fig. \ref{fig-fa}.    The fit has the duality constraint that  the vector and axial
  parts of  structure function  $W_{2}$ for quasielastic scattering are equal   
    at large $Q^2$.
 %
 %
 %
  
  The ratio of the  flux-weighted   MiniBooNE measured cross sections at low-$\nu$ to the calculated 
   (with the nominal published MiniBooNE flux) flux-weighted
   cross sections for any of the three models is
   proportional to the ratio of the `low-$\nu$' flux to the nominal MiniBooNE flux. As expected, the relative
   flux extracted as a function of neutrino energy is insensitive to the choice of  model. 
   
   The left side of Fig. \ref{fig-flux} shows the ratio of the flux extracted from  $\nu<0.1$ GeV events (red circles) and $\nu<0.2$ GeV events (blue squares) to the nominal flux.  Only statistical errors are shown. The green line is the systematic error in  nominal flux (as published by MiniBooNE). The extracted `low-$\nu$' flux is consistent with the nominal  flux within the quoted systematic errors.  The black curve (flux A) and purple curve (flux B) are possible deviations (which are consistent with the `low-$\nu$' flux)  from the nominal flux.
   
   Next we fit for the best value of $M_A$ for each of the three models. We find  that if
   we let the overall normalization float within the systematic error the extracted values of $M_A$
     using the  nominal flux,  flux A, and flux B  
     are within $0.015$ GeV of each other
as shown in Table 1.  We find that with Transverse Enhancement, and a modified dipole form factor,  the fit to the $Q^2$ dependence of the MiniBooNE $d\sigma/dQ^2$ on carbon  favors an axial mass $M_A=1.10\pm0.02$ GeV.  The ratio of this modified dipole  fit with $M_A=1.10$ GeV to the simple dipole parametrization with  $M_A=1.015$ GeV is shown as the solid red line in fig. \ref{fig-fa}.  The fit to the MiniBooNE data is more consistent with the values of $F_A(Q^2)$ extracted from pion electroproduction on free nucleons (shown in Fig. \ref{fig-fa}(b)), than with the values  $F_A(Q^2)$ extracted from neutrino data on deuterium (Fig. \ref{fig-fa}(a)).  
\begin{table}[ht]
\caption{\label{A:fits}  Fits to MiniBooNE neutrino quasielastic scattering data on carbon}
\begin{tabular}{|c|c|c|c|c|c|} \hline 
Form Factors &data~set & $M_A$ &$N$  &  $\chi^2/NDF$ &flux  \\ 
 vector/axial & (2/D)/(1D)   & (GeV)    & normalization &model\\   \hline
BBBA07               & double diff (2D)  & $1.35\pm 0.02$   &  $0.99 \pm 0.01$   & $39.8/ 135=0.30$  &nominal   \\ 
FA=Dipole   & $d\sigma/dQ^2$(1D)  &   $1.41 \pm 0.04$   &  $0.99 \pm 0.02$  & $11.7/15=0.78$   &nominal \\ \hline
BBC(TE)          & double diff  (2D)    &$1.22 \pm 0.02$   &  $1.01\pm 0.01$  & $43.6/135=0.32$  &nominal  \\ 
 FA=dipole  &  $d\sigma/dQ^2$(1D)  &   $1.17 \pm 0.03$   &  $1.05 \pm 0.02$  & $19.7/15=1.31$   &nominal  \\ \hline \hline
BBC(TE)     &  double diff (2D)         &  $1.17 \pm 0.02$    &  $1.01 \pm 0.01$  & $35.2/135=0.28$    &nominal  \\  FA=mod. dipole    &  $d\sigma/dQ^2$(1D)  &   $1.11 \pm 0.03$   &  $1.04 \pm 0.02$  & $19.0/15=1.27$    &nominal  \\ 
             \hline 
             BBC(TE)     &  double diff  (2D)         &  $1.17 \pm 0.02$    &  $1.01 \pm 0.01$  & $35.4/135=0.26$   &Flux A  \\  FA=mod. dipole    &  $d\sigma/dQ^2$(1D)  &   $1.10 \pm 0.03$   &  $1.04 \pm 0.02$  & $17.8/15=1.18$    &Flux A  \\   \hline
             BBC(TE)     &  double diff  (2D)          &  $1.17 \pm 0.02$    &  $1.01 \pm 0.01$  & $38.3/135=0.28$    &Flux B  \\  FA=mod. dipole   &  $d\sigma/dQ^2$(1D)  &   $1.09 \pm 0.03$   &  $1.04 \pm 0.02$  & $17.7/15=1.18$   &Flux B  \\
  \hline
\end{tabular}
\end{table}
%

\end{document}